\documentclass[reqno]{amsart}

\usepackage{booktabs}
\usepackage{IEEEtrantools}
\usepackage{amssymb,latexsym,amsfonts,amsmath}
\usepackage{mathrsfs}
\usepackage{graphicx}
\usepackage{url}
\usepackage{tikz}
\usetikzlibrary{calc,shapes,arrows}
\usepackage{amsfonts}

\usepackage{algorithm}
\usepackage{algorithmic}

\usepackage{xspace}

\newtheorem{theorem}{Theorem}[section]
\newtheorem{lemma}[theorem]{Lemma}

\newtheorem{problem}[theorem]{Problem}

\newtheorem{remark}[theorem]{Remark}
\newtheorem{assumption}{Assumption}
\numberwithin{equation}{section}

\usepackage{fancyhdr}

\newenvironment{nouppercase}{%
	\renewcommand{\uppercasenonmath}[1]{}}{}

\begin{document}

\begin{abstract}
In this paper, we introduce the notion of \emph{simulation-gap functions} to formally quantify the potential gap between an approximate nominal mathematical model and the high-fidelity simulator representation of a real system. Given a nominal mathematical model alongside a quantified simulation gap, the system can be conceptualized as one characterized by bounded states and input-dependent disturbances. This allows us to leverage the existing powerful model-based control algorithms effectively, ensuring the enforcement of desired specifications while guaranteeing a seamless transition from simulation to real-world application. To provide a formal guarantee for quantifying the simulation gap, we develop a data-driven approach. In particular, we collect data using high-fidelity simulators, leveraging recent advancements in Real-to-Sim transfer to ensure close alignment with reality. We demonstrate the effectiveness of the proposed method through experiments conducted on a nonlinear pendulum system and a nonlinear Turtlebot model in simulators. 
\end{abstract}

\title{{\LARGE Towards Mitigating Sim2Real Gaps: A Formal Quantitative\vspace{0.2cm}\\  Approach}$^*$\footnote[1]{$^*$This work was supported in part by the ARTPARK and the Siemens.}}

\author{{\bf {\large P Sangeerth}}$^1$}
\author{{\bf {\large Abolfazl Lavaei}}$^2$}
\author{{\bf {\large Pushpak Jagtap}}$^1$\\
	{\normalfont $^1$Indian Institute of Science, Bengaluru, India}\\
    {\normalfont $^2$School of Computing, Newcastle University, United Kingdom}\\
\texttt{\{sangeerthp,pushpak\}@iisc.ac.in}, \texttt{abolfazl.lavaei@newcastle.ac.uk}}

\pagestyle{fancy}
\lhead{}
\rhead{}
  \fancyhead[OL]{P Sangeerth, Abolfazl Lavaei, Pushpak Jagtap}

  \fancyhead[EL]{Towards Mitigating Sim2Real Gaps: A Formal Quantitative Approach} 
  \rhead{\thepage}
 \cfoot{}
 
\begin{nouppercase}
	\maketitle
\end{nouppercase}

\section{INTRODUCTION}
\label{introduction}
A plethora of literature explores the design of controllers based on the idea of having precise mathematical models \cite{Khalil:1173048}. Nevertheless, deriving an accurate mathematical model of the system is infeasible in most real-world applications. In such scenarios, the utility of learning-based and data-driven methods becomes evident (refer to \cite{jiang2020learning,brunton2022data,bazanella2011data,sutton2018reinforcement,tayal2024learning} and references therein), offering effective means to design controllers that enforce desired specification. It is noteworthy that those methods often fall short in providing formal guarantees to ensure satisfaction of the specified requirements. 

In recent years, several studies have introduced formal guarantees for learning-based and data-driven techniques. For stochastic systems, \cite{9095986,hasanbeig2019reinforcement} offer guarantees via reinforcement learning. Gaussian processes have been used for controller design with probabilistic guarantees \cite{jagtap2020control, 8909368, lederer2021gaussian}, but these methods assume boundedness of the RKHS norm of the kernel function. Other data-driven techniques \cite{9683211, akbarzadeh2024data,ROTULO2022110519, kenanian2019data,samari2024data,berger2021chance, 10149443, nejati2023formal,samari2024single} use control Lyapunov functions, control barrier functions, finite abstractions, and robust control techniques to provide deterministic or probabilistic guarantees. However, these methods often assume no model information is available, requiring extensive data collection.

While acquiring vast amounts of data from hardware is challenging, recent advances in sensor technology provide precise data collection solutions. These advancements have improved simulation software, reducing the gap between real-world and simulated environments \cite{chen2022real2sim,9363564}. Advanced simulators like LGSLV \cite{rong2020lgsvl}, ADAMS-Simulink \cite{brezina2011using}, Unreal physics engine \cite{unrealengine}, Metamoto \cite{samak2021autodrive}, NVIDIA’s Drive Constellation\footnotemark[3]{}, CarMaker\footnotemark[4]{}, and OpenDRIVE\footnotemark[5]{} enhance simulation fidelity using sensors, physics, and uncertainty models, providing valuable data for developing controllers with data-driven approaches.
\footnotetext[3]{https://resources.nvidia.com/en-us-auto-constellation/drive-constellation}
\footnotetext[4]{https://ipg-automotive.com/en/products-solutions/software/carmaker/}
\footnotetext[5]{https://www.asam.net/standards/detail/opendrive/}

Several studies, such as \cite{9606868,9993797,10229866,9723881,tan2018sim,akella2023safety}, concentrate on bridging the gap between simulation environments and reality within the Sim2Real context. Our paper aims to formally quantify this gap (\emph{i.e.,} sim2real gap) between the two environments. In \cite{weibel2021measuring}, the Sim2Real gap was measured for object classification. In reinforcement learning, it has been quantified as the difference in expected rewards between simulated and real systems \cite{tan2018sim}, but without formal guarantees. In \cite{akella2023safety}, the author quantifies the Sim2Real gap as a numerical value with a probabilistic guarantee. Our approach quantifies this gap as a function of states and inputs, with $100\%$ correctness guarantees.

Drawing on George Box's insight that ``All models are wrong, but some are useful'' \cite{doi:10.1080/01621459.1976.10480949}, our paper utilizes high-fidelity simulation software to gather data and presents a method employing an approximate yet effective mathematical model for designing controllers that seamlessly transition to real-world applications with formal guarantees. We adopt a data-driven framework to quantify the simulation-gap function between the nominal mathematical model and the high-fidelity simulator model, operating under the assumption that the latter closely approximates reality. Once this simulation gap is determined, we leverage existing model-based control techniques (\emph{e.g.,} \cite{Reissig_2017}) to synthesize a controller that operates seamlessly with the high-fidelity simulator model.  We demonstrate this claim with two physical case studies: a Pendulum (simulator model in Py-Bullet) and a Turtlebot (simulator model in Gazebo). Using the calculated simulation-gap functions for these two systems along with a nominal mathematical model, a symbolic controller is designed to enforce the desired specification in simulators. 

\section{System Description and Problem Definition}
\label{preliminaries section}

\textbf{Notations.} We denote sets of non-negative integers, positive real numbers, and non-negative real numbers, respectively, by $\mathbb{N}:=\{0,1,2,3,\dots\}$, $\mathbb{R}^+$, and $\mathbb{R}^+_0$. The absolute value of $x \in \mathbb{R}$ is denoted by $|x|$. The Euclidean norm of $x \in \mathbb{R}^n$ is represented by $\lVert x\rVert$. Symbol ${\mathbb{R}^n}$ is used to denote an $\textit{n}$-dimensional Euclidean space. A column vector $x \in \mathbb{R}^n$ is denoted by $x\hspace{-0.2em}=\hspace{-0.2em}[x_1;x_2;\cdots;x_n]$. For a matrix $A \in \mathbb{R}^{m \times n}$, $A^\top$ denotes its transpose.

\subsection{System Description}
\label{system description subsection}
In this paper, we consider a nominal mathematical model for the system represented by a discrete-time system, derived through the discretization of conventional continuous-time models \cite{rabbath2013discrete}, with a sampling time $\tau\in\mathbb{R}^+$, expressed as 
\begin{equation}
\label{disc_mathematical_model}
    {\Sigma}\!: x(k+1) = {f}_{\tau}(x(k),u(k)), \quad k \in \mathbb{N},
\end{equation}
where:
\begin{itemize}
    \item  $x =[x_1; x_2; \cdots ; x_n] \in X \subset \mathbb{R}^n$ is the state vector  in a bounded state set $X$;
    \item  $u=[u_1;u_2;\cdots; u_m] \in U\subset \mathbb{R}^m$ is the input vector in a \emph{finite} input set $U$ with cardinality $M$;
    \item ${f}_{\tau}:\textit{X} \times \textit{U} \rightarrow \textit{X}$ is the known transition map which is assumed to be locally Lipschitz continuous to guarantee the uniqueness and existence of the solution \cite{Khalil:1173048}.
\end{itemize}

\begin{remark}
In this work, motivated by real-world scenarios where the controller is deployed on a digital platform with quantized inputs, we have chosen to work with a finite input set. Nevertheless, our proposed approach can be also adapted to accommodate a continuous input set.
\end{remark}
While literature often focuses on designing controllers based on exact mathematical models, these controllers can fall short in practice. Recent advancements in sensor technology offer accurate real-state measurements. Consequently, there has been significant progress in the development of high-fidelity simulators like Gazebo \cite{1389727}, CARLA \cite{dosovitskiy2017carla}, Webots \cite{Webots}, Pybullet \cite{coumans2021}, and MuJoCo \cite{6386109} which closely replicate real systems using sensor models, physics engines, and uncertainty models. We assume that the real-world control system evolution in these simulators can be effectively represented by an unknown discretized map with the same discretization parameter $\tau$, as 
\begin{equation}\label{simulator_model}
    \hat{\Sigma}\!: x(k+1) =\hat{f}_{\tau}(x(k),u(k)), \quad k \in \mathbb{N},
\end{equation}
where $\hat{f}_{\tau}:\textit{X} \times {U} \rightarrow \textit{X}$ represents the transition map of the high-fidelity simulator model. For brevity, the $f_{\tau}$ and $\hat{f}_{\tau}$ will be represented as $f$ and $\hat{f}$ in the further discussions. 

\subsection{Problem Formulation}
\label{problem formulation section}

Now, we formally outline the problem under consideration in this work. 
\begin{problem}
    \label{Problem-1}
    Given the nominal mathematical model ${\Sigma}$ and high-fidelity simulator $\hat{\Sigma}$, we aim to
    \begin{enumerate}
        \item[(i)] formally quantify a simulation gap between the mathematical model and the high-fidelity simulator model, represented by a function $\gamma(x,u)\!:X \times U \rightarrow {\mathbb{R}^+_0}^n $ such that for all $x \in X,$ and $u \in U $,
        $\hat{f}(x,u) \in {f}(x,u)+[-\gamma(x,u),\gamma(x,u)]$;
        \item[(ii)] design a controller using the quantified simulation gap $\gamma(x,u)$ and the nominal mathematical model to meet specifications in the high-fidelity simulator ${\Sigma}$.
    \end{enumerate}
\end{problem}
To solve Problem~\ref{Problem-1}, we use data from both the mathematical model and high-fidelity simulator to derive the simulation gap function via a convex optimization problem. This allows us to design a controller for the mathematical model that meets specifications in the simulator.

\section{Data-driven Framework}\label{data-driven framework section}
In general, an unpreventable gap exists between the mathematical models and the models present in the high-fidelity simulators. The following lemma introduces the simulation-gap function, which quantifies this difference between the mathematical model $\Sigma$ and the high-fidelity simulator model $\hat{\Sigma}$.
\begin{lemma}\label{basic_bound_lemma}
    Consider the nominal mathematical model ${\Sigma}$ and its unknown simulator dynamics, denoted by $\hat{\Sigma}$.
If for all $i \in \{1,2,\dots, n\}$, there exists a function map $\gamma_i:X \times U \rightarrow \mathbb{R}_0^+$, for all $x \in X$, for all $ u \in U$, such that $|\hat{f_i}(x,u)-{f_i}(x,u)| \leq \gamma_i(x,u)$, then 
\begin{equation}
    \label{basic_bound_lemma_equation}
    \hat{f}(x,u) \in {f}(x,u)+[-\gamma(x,u),\gamma(x,u)],
\end{equation}
where $\gamma(x,u)=[\gamma_1(x,u);\gamma_2(x,u);\cdots;\gamma_n(x,u)],$ $f(x,u)=[f_1(x,u);f_2(x,u);\dots;f_n(x,u)],$ and $\hat{f}(x,u)=[\hat{f_1}(x,u);\hat{f_2}(x,u);\cdots;\hat{f_n}(x,u)].$
\end{lemma}

{\begin{proof}
The proof is straightforward. Let us assume for all $i \in \{1,2,\dots, n\}$, there exists an upper bound $\gamma_i(x,u) \geq 0$, for all $x \in X$, for all $ u \in U$, such that, we have $|\hat{f_i}(x,u)-{f_i}(x,u)| \leq \gamma_i(x,u)$. Then one has
${f_i}(x,u)-\gamma_i(x,u) \leq  \hat{f_i}(x,u) \leq {f_i}(x,u)+\gamma_i(x,u).$
Since, the above equation holds true for any $i \in \{1,2,\dots, n\}$,
\begin{equation*}
      \hat{f}(x,u) \in {f}(x,u)+[-\gamma(x,u),\gamma(x,u)]
\end{equation*}
holds true for all $x \in X$, for all $ u \in U$, which concludes the proof. 
\end{proof}}
We refer to $\gamma(x,u)$ as the simulation-gap function between $\hat{\Sigma}$ and ${\Sigma}$.

\begin{remark}
    Note that $\gamma_i(x,u)=\infty$ is a trivial upper bound for $|\hat{f_i}(x,u)-{f_i}(x,u)|$. In this work, our aim is to obtain the tightest upper bound by minimizing $\gamma_i(x,u)$ in our proposed optimization setting.
\end{remark}

The conditions mentioned in Lemma \ref{basic_bound_lemma}, for $i \in \{1,2,\dots,n\}$, can be reformulated as the following robust optimization program (ROP):
\begin{equation}
\label{eq:ROP_for_one_state}
\begin{aligned}
\min_{\gamma_i \in \mathcal{H},\eta_i \in \mathbb{R}}
&\eta_i  \\
\text{s.t.}\quad\quad &\gamma_i(x,u) \leq \eta_i, \quad \forall x\!\in\! X, \forall u \!\in\! U,\\
& |\hat{f_i}(x,u)-{f_i}(x,u)| \!-\! \gamma_i(x,u) \!\leq\! 0,~~\!\! \quad \forall x\!\in\! X, \forall u \!\in\! U,\\
\end{aligned}
\end{equation}
where $\mathcal{H}:=\{g \,\big|\, g\!:X \times U \rightarrow \mathbb{R}\}$ is a functional space. 

One can readily observe considerable challenges in solving the ROP in \eqref{eq:ROP_for_one_state}. Firstly, the ROP in \eqref{eq:ROP_for_one_state} has infinite constraints since the state space is continuous. Secondly, the function map $\hat{f_i}(x,u)$ is unknown.
To tackle these challenges, we aim to develop a data-driven scheme proposed in \cite{10149443}, for computing the simulation gap function without directly solving the ROP in \eqref{eq:ROP_for_one_state}. To achieve this, we initially gather a set of $N$ sampled data points within $X$ by considering balls $X_r$ around each sample $x_r$, $r\in\{1,\ldots,N\}$ with radius $\epsilon$ such that $X \subseteq \bigcup_{r=1}^NX_r$ and 
\begin{equation}
\label{max_epsilon}
    \Vert x-x_r \Vert \leq \epsilon, \quad \forall x \in X_r.
\end{equation}
Recall that the input space is already finite with cardinality $M$. Using these $N$-representative points $x_r \in X_r$ as initial conditions and for all the $M$-input points $u \in U$, we collect $N \times M$ data from the mathematical model and the high-fidelity simulator for the next sampling instance $\forall i \in \{1,2,\ldots,n\}$, which is represented as
 \begin{align}
 \label{data_eqn}
    \big\{\hspace{-0.1em}(x_r,u,\hat{f_i}(x_r,u),{f_i}(x_r,u))\,\big|\, r\in\{1,\ldots,N\}, \forall u \in U\big\}.
\end{align}
Now, we select the structure of $\gamma_i(x,u)$ to render the optimization program convex with respect to decision variables. In our data-driven setting, we assume that the structure of $\gamma_i(x,u)$ can be fixed as $\sum_{l=1}^{z_i} q^{(l)}_ip^{(l)}_i(x,u)$, a parametric form linear in decision variables $q_i= [q^{(1)}_i;\cdots;q^{(z_i)}_i] \in \mathbb{R}^{z_i}$, with (potentially nonlinear) user-defined basis functions $p^{(l)}_i(x,u)$. The basis functions $p^{(l)}_i(x,u)$ can be considered to have any arbitrary form. This renders our optimization program convex in nature.
The ROP described by \eqref{eq:ROP_for_one_state} for all $i \in \{1,2,\dots,n\}$, is now reformulated as the following scenario convex program (SCP) based on data:
\begin{align}
\min_{q_i \in \mathbb{R}^{z_i}, \eta_i \in \mathbb{R}} &\eta_i  \nonumber\\
\text{s.t.}\quad & q_i^\top p_i(x_r,u) \leq \eta_i, \quad \forall r\in\{1,\ldots,N\}, \forall u \!\in\! U, \nonumber\\
& |\hat{f_i}(x_r,u)-{f_i}(x_r,u)|-q_i^\top p_i(x_r,u)\leq 0, \nonumber\\&\forall r\in\{1,\ldots,N\}, \forall u \in U.\label{eq:SOP_for_one_state}
\end{align}
We aim to compute the optimal solution $\eta^{SCP}_i$ to obtain the minimum value of ${q_i}^\top p_i(x_r,u),$ which will subsequently be utilized to upper bound the simulation-gap function.
\section{Formal Quantification of Simulation Gap}
\label{quantification of simulation gap section}
To solve the constructed SCP in~\eqref{eq:SOP_for_one_state}, we first raise the following assumption.

\begin{assumption}\label{A1}
 Let the functions $|\hat{f_i}(x,u)-{f_i}(x,u)|$ and ${q_i}^\top p_i(x,u)$ be Lipschitz continuous with respect to $x$ with Lipschitz constants $\mathcal{L}^{(i)}_1$ and $\mathcal{L}^{(i)}_2$, for any finite $u \in U$.
\end{assumption}

Even though the function $\hat{f_i}(x,u)$ is not known, one can employ the proposed results in \cite{Wood1996EstimationOT} or Algorithm-1 in \cite{10149443} to estimate the Lipschitz constants of both the functions mentioned in Assumption~\ref{A1}, using a finite number of data collected from unknown systems. Under Assumption~\ref{A1}, we now present the following theorem, as the main result of the work, to compute the actual upper bound for the $|\hat{f_i}(x,u)-{f_i}(x,u)|$. 
\begin{theorem} \label{Theorem 2}
Let Assumption \ref{A1} hold. Consider the solution $q_i=[q_i^{(1)};\cdots;q_i^{(l)}]$ with an optimal value $\eta_i^{SCP}$ of SCP in \eqref{eq:SOP_for_one_state}. Then for all $i \in \{1,2,\dots,n\}$, the absolute difference between the state evolutions of $\Sigma$ and $\hat\Sigma$ is quantified as:
\begin{equation}
\label{actual_gamma}
    |\hat{f_i}(x,u)-{f_i}(x,u)| \leq {q_i}^\top p_i(x,u) + \mathcal{L}^{(i)} \epsilon, 
\end{equation}
for all $x \in X$, for all $u \in U$, where the constant $\mathcal{L}^{(i)}=\mathcal{L}^{(i)}_1+ \mathcal{L}^{(i)}_2$ and $\epsilon$ as in \eqref{max_epsilon}.
\end{theorem}
{\begin{proof}For a given $i\in\{1,2,\dots,n\}$, for all $x\in X$ and for all $u\in U$, we have
\begin{align*}
|\hat{f_i}(x,u)\hspace{-0.2em}-\hspace{-0.2em}{f_i}(x,u)|=~\!&|\hat{f_i}(x,u)\hspace{-0.2em}-\hspace{-0.2em}{f_i}(x,u)|\hspace{-0.2em}-\hspace{-0.2em}|\hat{f_i}(x_r,u)\hspace{-0.2em}-\hspace{-0.2em}{f_i}(x_r,u)|+|\hat{f_i}(x_r,u)-{f_i}(x_r,u)|,
\end{align*}
where $x_r\in X_r$ such that $\|x-x_r\|\leq\epsilon$. Under Assumption \ref{A1} and $q_i^\top p_i(x_r,u)$ obtained by solving SCP in \eqref{eq:SOP_for_one_state}, one can obtain the following series of inequalities:
\begin{align*}
|\hat{f_i}(x,u)&-{f_i}(x,u)|\\
&\leq \mathcal{L}^{(i)}_1 \Vert x-x_r \Vert +|\hat{f_i}(x_r,u)-{f_i}(x_r,u)|\\
&\leq \mathcal{L}^{(i)}_1 \epsilon + {q_i}^\top p_i(x_r,u)\\
&= \mathcal{L}^{(i)}_1 \epsilon + {q_i}^\top p_i(x_r,u)-{q_i}^\top p_i(x,u)+{q_i}^\top p_i(x,u)\\
&\leq \mathcal{L}^{(i)}_1 \epsilon +\mathcal{L}^{(i)}_2 \Vert x-x_r \Vert +{q_i}^\top p_i(x,u)\\
&\leq \mathcal{L}^{(i)} \epsilon +{q_i}^\top p_i(x,u),
\end{align*}
which concludes the proof.
\end{proof}}
By following Lemma \ref{basic_bound_lemma} and equation \eqref{actual_gamma}, we quantify the simulation-gap function for the entire state-space as $\gamma_i(x,u):={q_i}^\top p_i(x,u) + \mathcal{L}^{(i)} \epsilon$. 
\begin{remark}
    Note that for a system of order $n$, one needs to solve $n$ SCPs to find $\gamma_i(x,u)$, for all $i \in \{1,2,\dots,n\}$.
\end{remark}

\begin{remark}\label{epsilon_gamma_remark}
    One can readily observe from \eqref{actual_gamma} that if a smaller value of $\epsilon$ is chosen, one can get a tighter bound for $\gamma_i(x,u)$. This is further illustrated in Table \ref{Pendulum epsilon table}. However, this can lead to an increase in computational complexity.
\end{remark}

\section{Controller Synthesis}
\label{controller synthesis section}
We now design a controller for the high-fidelity simulator system $\hat{\Sigma}$, where the dynamics is now represented with the help of nominal mathematical model \eqref{disc_mathematical_model} and the formally quantified simulation gap \eqref{actual_gamma} as
\begin{equation}
    \label{basic_bound_equation}
    x(k+1) \in {f}(x(k),u(k))+[-\gamma(x(k),u(k)),\gamma(x(k),u(k))].
\end{equation}
This system can be perceived as one with bounded disturbance, 
\begin{equation*}
    \label{basic_bound_equation_basic}
    x(k+1) \in {f}(x(k),u(k))+[-w,w],
\end{equation*}
a topic extensively covered in existing literature on controller design for systems with uncertainty. Examples include model predictive control (MPC) \cite{MAYNE2005219}, adaptive control \cite{singh2022adaptive}, and symbolic control \cite{tabuada2009verification}. In our work, we have employed symbolic controllers \cite{Reissig_2017} for our case studies, utilizing the SCOTS toolbox \cite{rungger2016scots}. The implementation details of the controller are not provided due to space constraints, and readers are encouraged to refer to \cite{Reissig_2017,rungger2016scots} for more details.

\section{Case Studies}
\label{case studies subsection}
To show the effectiveness of our results, we demonstrate our data-driven approach to a \emph{pendulum system} and a \emph{unicycle model of Turtlebot}, both with nonlinear dynamics. We used a computer with AMD Ryzen 9 5950x, 128 GB RAM, and an NVIDIA RTX 3080Ti graphics card to collect data. For the mathematical model, the data was collected from MATLAB for both case studies. 
In the Pendulum example, the high-fidelity simulator model data was collected from PyBullet, while for the Turtlebot example, the high-fidelity simulator model data was collected from Gazebo.
\subsection{Pendulum System}
\label{pendulum}
As the first case study, we consider a pendulum system whose mathematical model is given as follows:
\begin{figure}
\centering
  \includegraphics[scale=0.28]{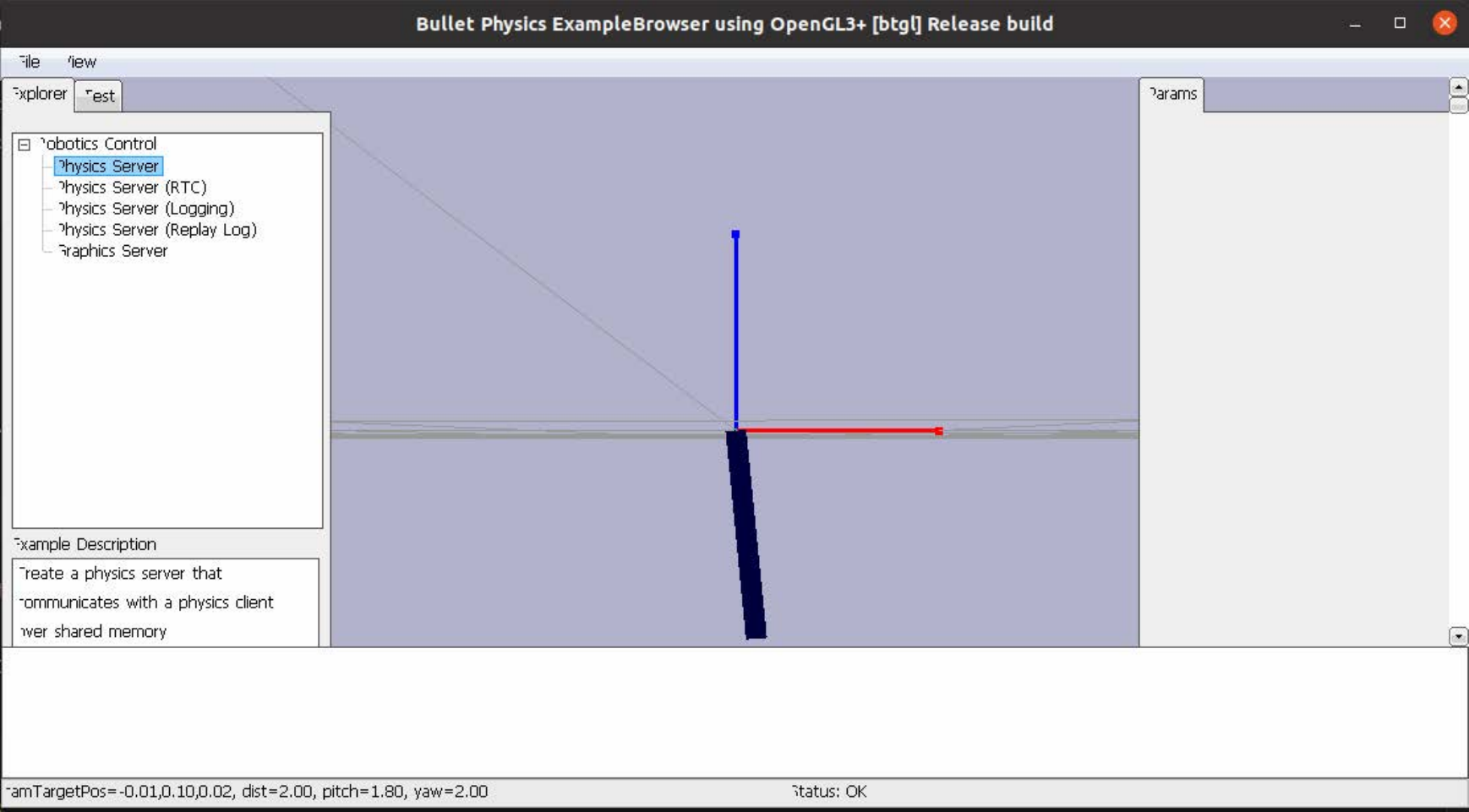}
  \caption{Pendulum model in the Py-Bullet simulator.}
  \label{fig:Pendulum model in Py-Bullet simulator}
\end{figure}
\begin{equation*}
    \begin{bmatrix}
    x_1(k+1)\\
    x_2(k+1)
\end{bmatrix}=
\begin{bmatrix}
    x_1(k)+\tau x_2(k)\\
    -\frac{3g\tau}{2l}\sin{x_1(k)}+x_2(k)+\frac{3\tau u(k)}{ml^2}\\
\end{bmatrix}\!\!,
\end{equation*}
where $x_1$, $x_2$, and $u$ are the angular position, angular velocity, and torque input, respectively. The parameters $m=1kg$, $g=9.81m/s^2$, and $l=1m$ are, respectively, the pendulum's mass, acceleration due to gravity, and rod length. The pendulum model is simulated in the high-fidelity simulator PyBullet, depicted in Fig. \ref{fig:Pendulum model in Py-Bullet simulator}. The parameter $\tau$ is the sampling time chosen as $0.005s$. The state-space is $X=[-0.2,0.2] \times[-0.5,0.5]$, while the finite input space is considered $U=\{-1.2,-1.1,\dots,1.1,1.2\}$. 
The data is collected with the state-space discretization $\epsilon=0.0022$. After collecting data, we fix the structure of $\gamma_i(x,u)$ as $q^{(1)}_{i}x_1^2+q^{(2)}_{i}x_2^2+q^{(3)}_{i}x_1x_2+q^{(4)}_{i}x_1+q^{(5)}_{i}x_2+q^{(6)}_{i}u+q^{(7)}_{i}$ for all $i \in \{1,2\}$. We solve SCP in \eqref{eq:SOP_for_one_state} and obtain $q^{(2)}_{1}=0.0012$, $q^{(3)}_{1}=0.0003$, $q^{(7)}_{1}=0.0008$, $q^{(1)}_{2}=-0.5657$, $q^{(2)}_{2}=-0.0139$, $q^{(3)}_{2}=-0.0482$, $q^{(4)}_{2}=0.1809$, $q^{(5)}_{2}=0.0022$, $q^{(7)}_{2}=0.0081$, and $q^{(1)}_{1}=q^{(4)}_{1}=q^{(5)}_{1}=q^{(6)}_{1}=q^{(6)}_{2}\approx 0$ with optimal values $\eta_1^{SCP}=0.0012$ and $\eta_2^{SCP}=0.0245$, respectively. With $\epsilon=0.0022$, the total time to collect data and solve the optimization problem was around 1.9 hours. Terms with coefficients, as determined by the solver, of order lesser than $10^{-6}$ are neglected in the further representations.

\begin{figure}[!ht]
    \centering
    \hspace{-1.9em}
    \includegraphics[scale=0.19]{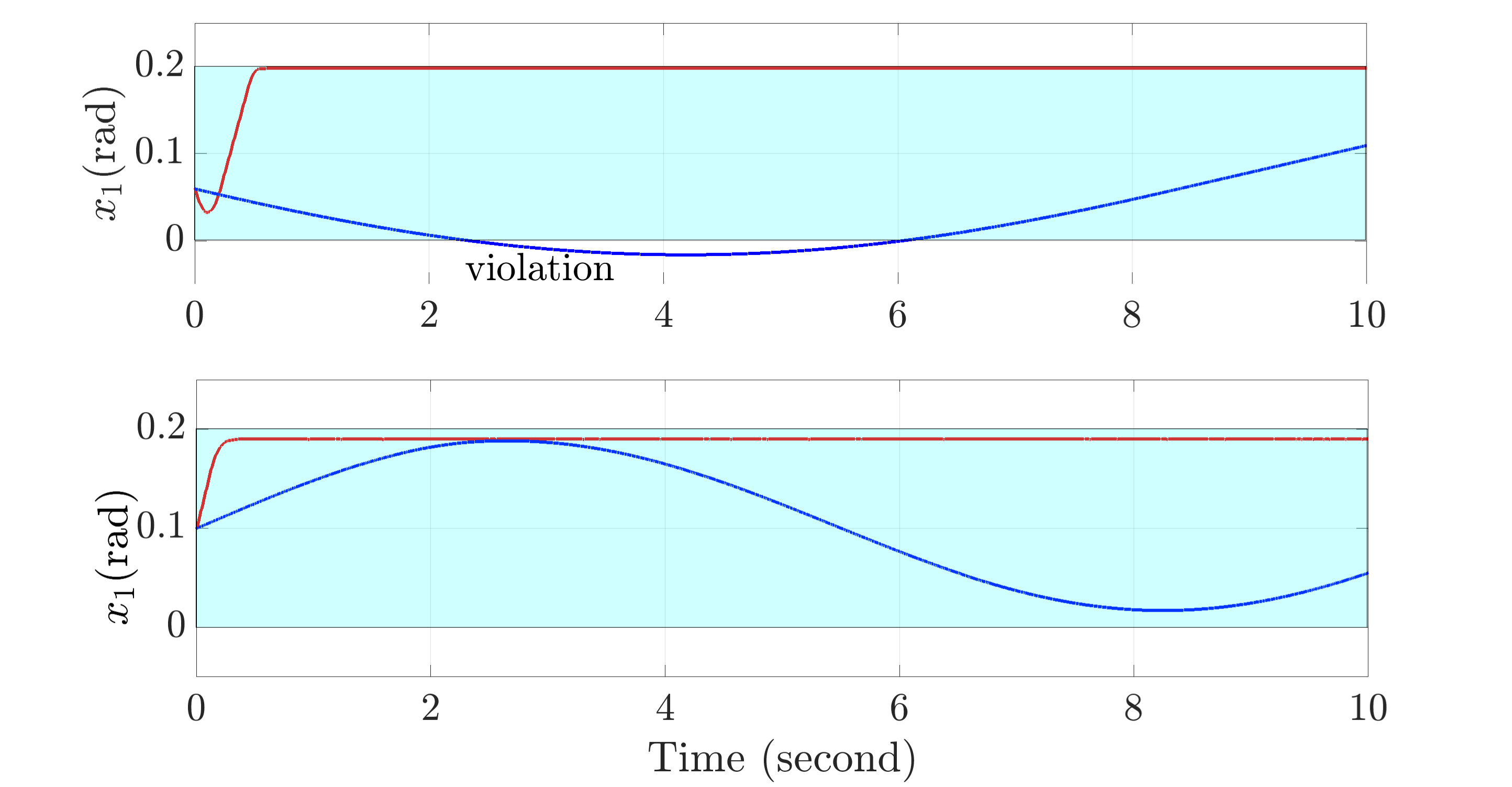}
    \caption{
    The state trajectory $x_1$ for both the mathematical system (red) and the PyBullet model (blue) is shown. The invariance specification is violated in PyBullet when the controller is synthesized without considering $\gamma(x,u)$ (top). The state-space invariance specification is satisfied in PyBullet when the controller is synthesized considering $\gamma(x,u)$ (bottom).}
    \label{fig:combined pendulum result}
\end{figure}

To compute a simulation gap $\gamma_i(x,u)$ for the entire state-space with guarantees, we compute the Lipschitz constants using the Algorithm-1 of \cite{10149443} as $\mathcal{L}_1^{(1)}=1.2087,$ $\mathcal{L}_2^{(1)}=0.0013$, $\mathcal{L}_1^{(2)}=7.2331$, and $\mathcal{L}_2^{(2)}=0.1680$.
Following \eqref{actual_gamma}, the simulation gap functions are quantified as 
\begin{align*}
\gamma_1(x,u)&=0.0012x_2^2-0.0003x_1x_2+0.00756,\\
\gamma_2(x,u)&=-0.5657x_1^2-0.0139x_2^2-0.0482x_1x_2+0.1809x_1-0.0022x_2+0.0495.
\end{align*}
Table \ref{Pendulum epsilon table} illustrates the impact of  $\epsilon$ on the value of the simulation gap for each dimension. It is evident from Table \ref{Pendulum epsilon table} that as the $\epsilon$ value decreases, the maximum value of $\gamma_i(x,u)$ also decreases, indicating that the simulation gap is becoming tighter. We now employ a symbolic control approach to synthesize controllers that enforce an invariance property (\emph{i.e.,} ensuring that the state remains within $[0,0.2]\times[-0.5,0.5]$). These controllers are synthesized using the symbolic controller toolbox SCOTS \cite{rungger2016scots} for both the nominal mathematical model and the model incorporating the simulation gap in \eqref{basic_bound_equation}. Fig. \ref{fig:combined pendulum result} depicts the trajectories of the mathematical model (red) and the PyBullet model (blue) using controllers designed from the nominal mathematical model (top) and the model with the quantified simulation gap as in \eqref{basic_bound_equation}(bottom). It is evident that the controller derived from the nominal mathematical model results in a violation of the invariance property when used for the PyBullet environment.  

\begin{table}[!t]
	\centering
	\caption{Variation of simulation gap with $\epsilon$}
	\begin{tabular}{|l|l|l|l|}
		\hline
		\hspace{3em} $\epsilon$ & 0.0110 & 0.0056 & 0.0022 \\ \hline
		$\sup\limits_{x\in X,u \in U} \gamma_1(x,u)$ &   0.0172   & 0.00757   & 0.00478    \\ \hline
		$\sup\limits_{x\in X,u \in U} \gamma_2(x,u)$ &   0.2028   &    0.0659   &    0.0555     \\ \hline
	\end{tabular}
	\label{Pendulum epsilon table}
\end{table}

\begin{figure}[!ht]
    \centering
    \hspace{-2em}
    \includegraphics[scale=0.189]{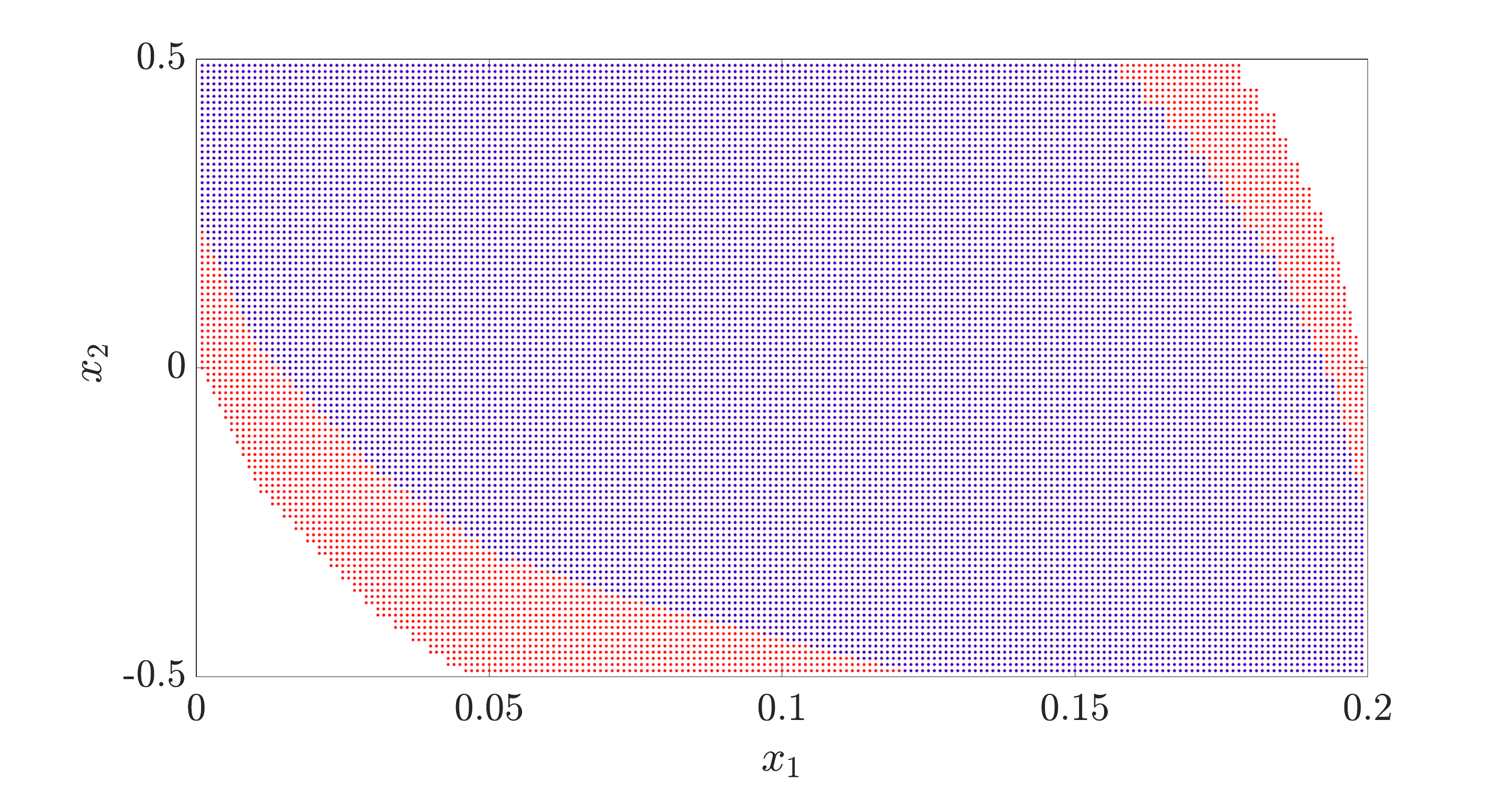}
    \caption{
    The red points represent the \emph{region of invariance} obtained by synthesizing the controller for the mathematical model without $\gamma(x,u)$. The blue points represent the region of invariance obtained by synthesizing after incorporating $\gamma(x,u)$ into the mathematical model.}
    \label{fig:winning domain pendulum}
\end{figure}
\begin{figure}
    \centering
    \hspace{-.2em}\includegraphics[scale=0.114]{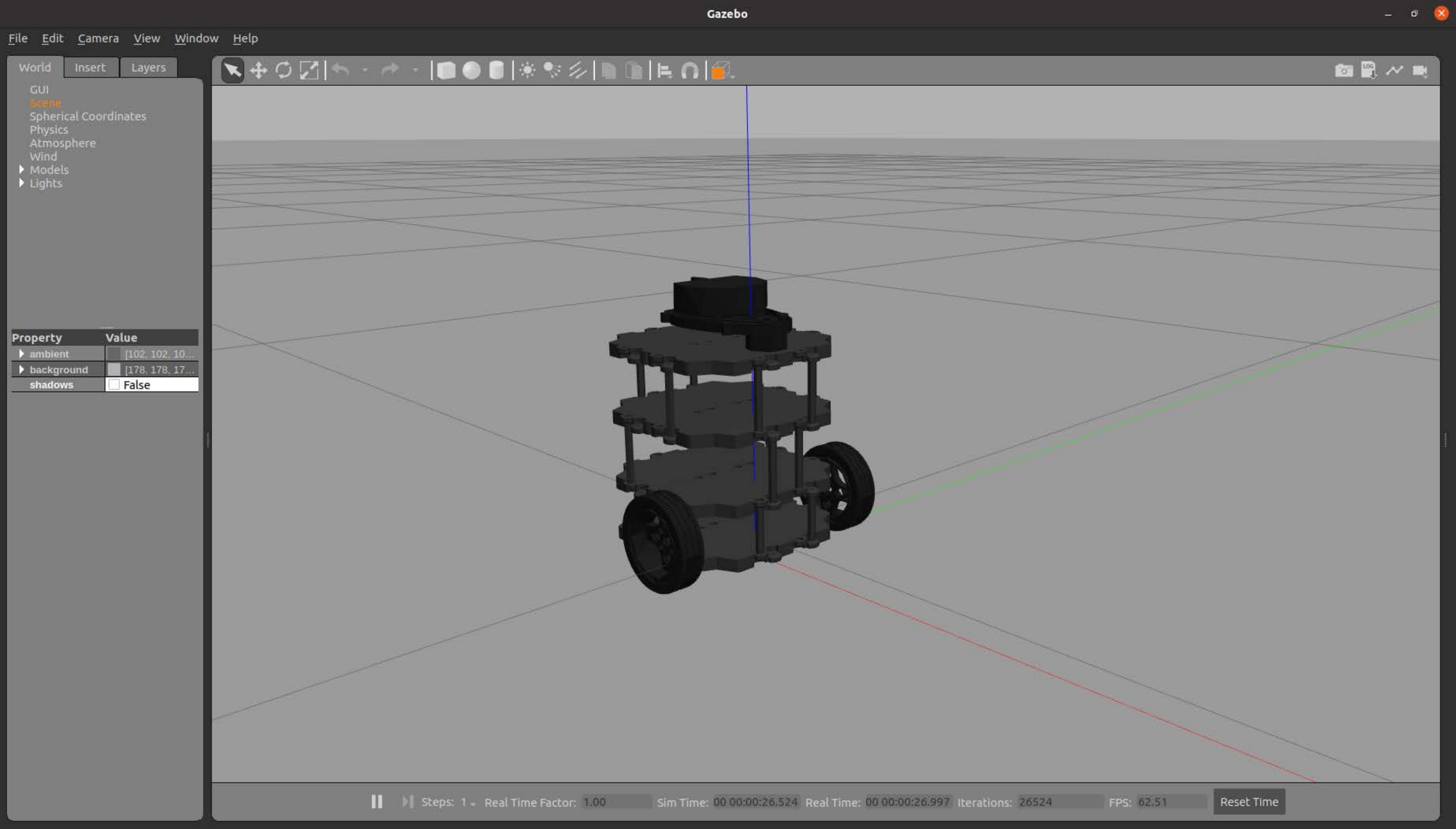}
    \caption{Turtlebot model in the Gazebo simulator.}
    \label{fig:Turtlebot pybullet}
\end{figure}

In Fig. \ref{fig:combined pendulum result}, two trajectories starting from different initial conditions are compared. Without $\gamma(x,u)$, the controller for the mathematical model yields the red invariant set in Fig. \ref{fig:winning domain pendulum}. However, applying the same controller to the simulator model results in a violation, as seen in the first plot of Fig. \ref{fig:combined pendulum result}. Adding the simulation gap function $\gamma(x,u)$ to the mathematical model produces a controller with a smaller, blue invariant set, shown in Fig. \ref{fig:winning domain pendulum}. This reduction indicates a gap between the original mathematical model and the simulator model, which prevents the simulator from satisfying the invariance property in the red zone. The figures use different initial conditions to highlight this gap. If initial conditions were in the intersection of the red and blue regions, the invariance property would be satisfied for both models mentioned in \eqref{disc_mathematical_model} and \eqref{basic_bound_equation}, obscuring the concept of the simulation gap function.

\subsection{Turtlebot Model}
\label{acceleration controlled unicycle model subsection}
For the second case study, we consider a unicycle model for a Turtlebot as the following: 
\begin{align*}
    \begin{bmatrix}
    x_1(k+1)\\
    x_2(k+1)\\
    x_3(k+1)\\
\end{bmatrix}=
\begin{bmatrix}
    x_1(k)+\tau u_1(k)\cos{x_3(k)}\\
    x_2(k)+\tau u_1(k)\sin{x_3(k)}\\
    x_3(k)+\tau u_2(k)\\
\end{bmatrix}\!,
\end{align*}
where $x_1$, $x_2$, $x_3$ denotes the position of the Turtlebot in the x and y axes and the orientation of the bot, respectively. The parameters $u_1$ and $u_2$ denote the linear and angular input velocities of the bot, respectively. The sampling time is chosen as $0.01s$. The Turtlebot model in a Gazebo simulator is shown in Fig. \ref{fig:Turtlebot pybullet}. The state and finite input sets are, respectively, considered as $X=[0,10]\times[0,10]\times[-\pi ,\pi]$ and $U=\{-1,-0.9,\dots,0.9,1\} \times \{-1,-0.9,\dots,0.9,1\}$.

\begin{figure}
    \centering
    \includegraphics[width=0.7\linewidth]{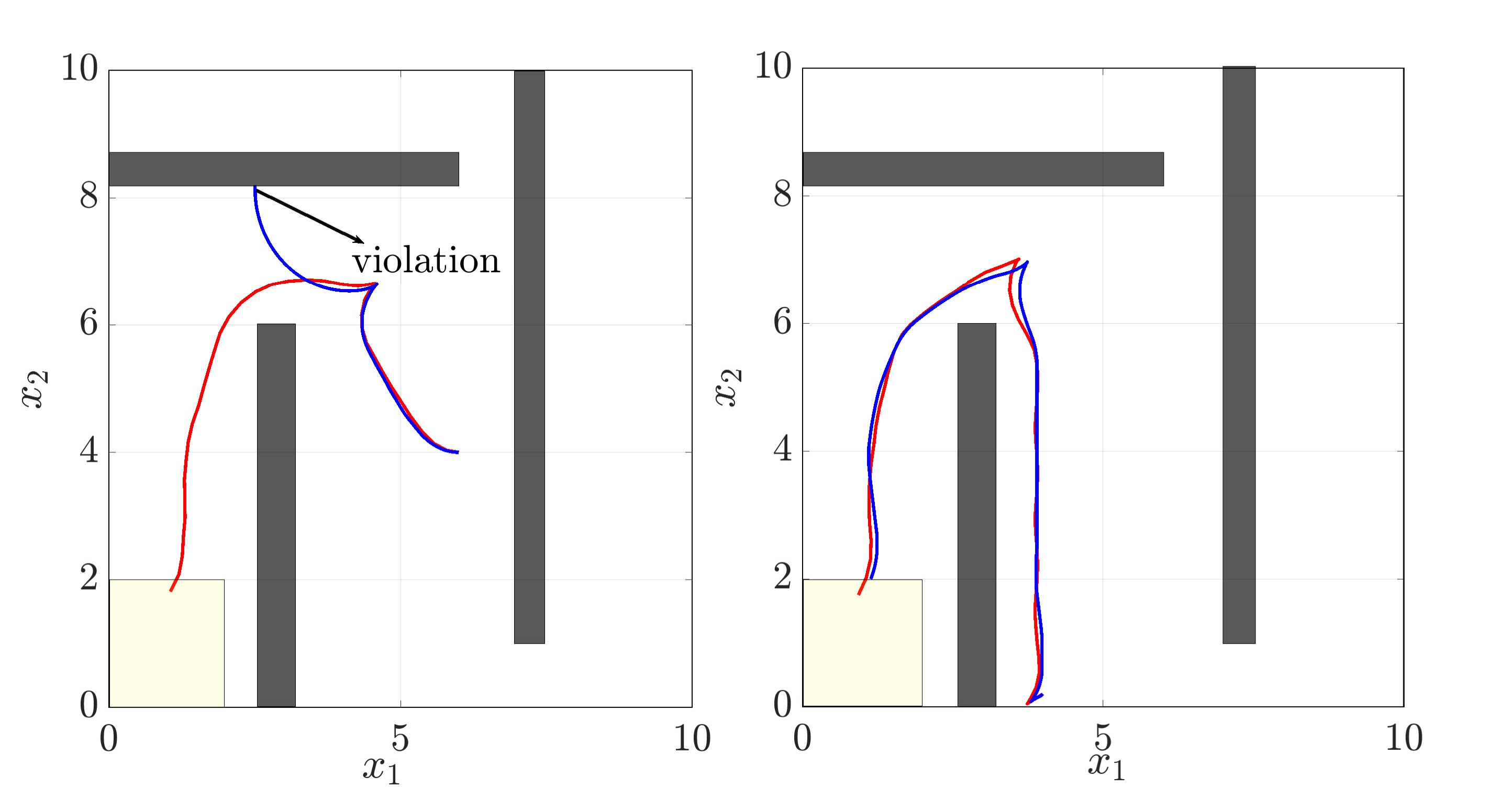}
    \centering
    \caption{
    State trajectories of both the mathematical model (red) and the Gazebo model (blue). The black regions represent the obstacles, while the yellow one represents the target. When the controller is synthesized for the reach-while-avoid specification without incorporating $\gamma(x,u)$, the Pybullet model hits the obstacles (left). The underlying specification is satisfied when the controller is synthesized after incorporating $\gamma(x,u)$ (right). }
    \label{fig:Turtlebot}
\end{figure}
The data is collected with $\epsilon=0.0087$. The structure of $\gamma_i(x,u)$ was fixed as $q_i^{(1)}x_1+q_i^{(2)}x_2+q_i^{(3)}+q_i^{(4)}u_1+q_i^{(5)}u_2$, $i\in \{1,2,3\}
$. Following the similar procedure discussed in the previous case study, we obtain the simulation gap functions as 
\begin{align*}
&\gamma_1(x,u)=0.0038u_1-0.0028u_2+0.0632,\\ 
&\gamma_2(x,u)=0.0058u_1-0.0004u_2+0.0632,\\ 
&\gamma_3(x,u)=0.0031u_1-0.0035u_2+0.2059.
\end{align*}
Terms with coefficients, as determined by the solver, of order lesser than $10^{-6}$ are neglected in the above representations. It is worth noting that in the Turtlebot example, the term $\gamma(x,u)$ depends more on the input than on the state, as the system’s \emph{continuous-time} kinematics model was driftless. In contrast, the pendulum system was not driftless, and the SCP results indicated that the physics engine of the PyBullet simulator required correction terms that depended more on the system states than on the input. The total time taken for data collection and solving the optimization problems for the Turtlebot example for the value of $\epsilon=0.0087$ was 7.3 hours. 

Using the SCOTS toolbox \cite{rungger2016scots}, we now design a symbolic controller for the mathematical model to satisfy the reach-while-avoid specification. When this controller, developed for the math model, is implemented directly into the simulator model, this controller does not fulfill the desired specification, and the robot hits the obstacle, as shown in Fig. \ref{fig:Turtlebot} (left). After adding simulation-gap function $\gamma(x,u)$ to the mathematical model dynamics, a symbolic controller is again synthesized for the same reach-while-avoid specification. One can readily see from Fig. \ref{fig:Turtlebot} (right) that the desired specification is now satisfied. It is worth noting that the mathematical model considered is itself a simple kinematic model, which does not consider many parameters such as mass, friction, and moment of inertia. Utilizing the proposed approach, one can design a controller feasible for a precise model (Gazebo Simulator) using the simple (inaccurate) mathematical model and the quantified simulation gap.

\section{Conclusions}
In this article, we presented a formal method for quantifying the gap between the nominal mathematical model and the high-fidelity simulator model. Our approach involves formally obtaining a state and input-dependent simulation gap function. If the high-fidelity simulator closely resembles the real-world environment, the proposed method facilitates the implementation of controller designed by incorporating the simulation gap function into the mathematical model, enabling seamless deployment of the same controller in real-world applications. We demonstrated our data-driven results over two physical case studies with nonlinear dynamics.

\bibliographystyle{alpha}
\bibliography{biblio}

\end{document}